# A Pilot Study Exploring Spreadsheet Risk in Scientific Research

Ghada AlTarawneh, Simon Thorne
Cardiff Metropolitan University

**Abstract**

This paper discusses the risks and potential impacts of spreadsheet errors in scientific research data in a Neuroscience research centre in the UK.

Spreadsheets usage in neuroscience, or indeed any medical discipline, is a largely unreported area of spreadsheet research. This paper presents a case study exploring the possible risks and impacts of spreadsheet errors in the neuroscience research centre at the University of Newcastle. Data was collected using an online questionnaire with 17 participants and two detailed semi-structured interviews.

The analysis highlights that errors in research data may lead to severe impacts such as misleading science and damaged personal and organisational reputations. In addition, many risks factors arise from using spreadsheets such as inadequate design and a lack of training.

Spreadsheets are used widely in business and the impacts and risks in these fields have been studied and highlighted in detail. However, scientific research and spreadsheets have also a significant relationship that has not been clarified. The paper also draws out the similarities in spreadsheet practice between the scientific and business communities.

## 1.0 Introduction

This paper discusses the risks and potential impacts of spreadsheet errors in scientific research data in a Neuroscience research centre in the UK.

Spreadsheets usage in neuroscience, or indeed any medical discipline, is a largely unreported area of spreadsheet research. Although little is published on this subject, it seems likely that the medical discipline will make extensive use of spreadsheets for a variety of clinical and non-clinical activities. This assumption is based on the observation that spreadsheet use is ubiquitous in almost all areas of business, government and education. To that end, this paper aims to answer the following questions

1. To what extent are spreadsheets used by the Neuroscience Research Centre at the University of Newcastle for data processing and decision making activities?
2. How are spreadsheets planned, developed and maintained by the research centre?
3. What are the specific risks and potential sources of error arising from Neuroscience spreadsheet use?
4. What is the likely impact of spreadsheet risks and error on neuroscience research data?

Data was collected using an online questionnaire with 17 participants and followed up by two detailed semi-structured interviews. Given the small sample size, only limited generalisations to the wider neuroscience and medical communities can be made. However, this research will gather some specific, interesting data and will highlight important areas for further research.

## 1.1 Neuroscience

Neuroscience was defined by the U.S. congress as *"the study of the nervous system, how it affects behaviour, and how it is affected by disease. The goal of neuroscience is to define and understand the continuum from molecular to cell to behaviour"* (Congress, 1984).

Several biological and human cognitive developments have been discovered in neuroscience research. Furthermore, many medical and mental health issues have been treated by its findings. Neuroscience did not gain the appropriate attention as a new science until the last decade.

In fact, studying the brain and the nervous system sets the foundation for many other studies. Psychology for example is a wide and rich discipline and many recent neuroscience research studies and experiments on the brain have explained many irrational attitudes and behaviours through studying precursor neural circuit activity (Diamond and Amso, 2008). Moreover, the impact of neuroscience on the wider medical discipline has been significant. Neuroscience research has made important contributions to the understanding of many medical conditions and has highlighted new avenues of research in multiple medical disciplines (Conn, 2008).

**1.2 Research Data confidentiality and Integrity**

As with other areas of research, confidentiality and data protection are of the upmost importance in neuroscience. There are several sets of standards and recommendations to that end such as the guidelines published by The National Human Research Protections Advisory Committee (NHRPAC) (NHRPAC, 2002). These guidelines highlight the importance of research data and therefore, the risks associated with it. However, there are no explicit guidelines for spreadsheets.

Research data services at Wisconsin-Madison University (WISC, 2014) discuss spreadsheets risks, errors and research data. They also published a set of guidelines and recommendations when using spreadsheets in order to minimize these errors in research data. Although these guidelines offer some basic advice, they are far from detailed.

- Research data in the organization is very important and losing it could lead to many results such as:
    - Losing valuable experimental data or simulation results.
    - Drawing incorrect conclusions from data leading to negative research impact.
    - Damage to the reputation of the organization conducting the research.
    - Potential for regulatory or legal action from governmental or other institution.
    - Requirement for corrective actions or repairs which would take years of research.
    - Violation of University or organization mission, policy, or principles.

This lack of detailed advice is in contrast to the regulation and control of spreadsheet applications in other medical fields such as the pharmaceutical industry in the United States. The pharmaceutical industry has highly specific controls placed on the use of spreadsheet applications for data analysis, reporting and decision making under Title 21 of the Code of Federal Regulations Part 11 (21 CFR 11). Spreadsheets are explicitly mentioned in this legislation which demands companies provide evidence of: audits, validation, electronic signatures and documentation for any software artefact including spreadsheets. The legislation also dictates that electronic artefacts such as spreadsheets be stored in a secure server so that once the spreadsheet has been created and audited, it cannot be changed without authorisation.

**1.3 Neuroscience data management strategies**

There are many approaches available to manage and distribute research data. Spreadsheets are one of the most popular data manipulation and analysis tools used among researchers.

Lacroix and Critchlow (2003) discuss spreadsheets as one of the two popular data management strategies in research. According to this research, spreadsheets offer quick data browsing, simple mathematical operations and easy distribution to collaborators. However, they also highlighted many points as disadvantages, in particular the lack of data validation when entered which can increase the possibility of errors.

In their research (Anderson, et al., 2007), Anderson and his colleagues interviewed 286 researchers from different research fields including neuroscience. The majority of the researchers interviewed admitted that they rely on general-purpose applications such as spreadsheets to manage their data. The main reason for this reliance is the simplicity of interface, the range and power of data manipulation tools and its short learning curve. According to one of the interviewees *"Yeah, the spreadsheet has been our main workhorse, unfortunately"*.

Anderson et al. (2007) also note that several of the interviewees have encountered problems when using spreadsheets:

> *"Well, we have multiple spreadsheets - that's one of the problems. We sort of have a master spreadsheet ... We try to minimize it as much as we can, but I think that's a major problem."*
>
> *"However, that exceeds the capabilities of the spreadsheet. Spreadsheet really bogs down any time you get past say 20,000 individual cells with columns."*
>
> *"Well, it's very cumbersome, I can't print anything, I'd have to paste it together. I end up just doing a freeze frame so that I can scroll this way."*

Although Anderson et al (2007) do not mention the wealth of research on spreadsheet error, it is clear from the quotes on spreadsheet problems that users experience problems with concurrency, computational power and usability.

Spreadsheets are widely used in research for tabulating, analysing and sharing data; *"Recent research in multiple disciplines shows that the use of spreadsheets to store and structure numeric and text data is commonplace"* (WISC, 2014). The main reason for this is probably the same as the reason that business rely on spreadsheets. It represents a fast way to test hypotheses, plot data, conduct pilot experiments and prototype ideas. Moreover, it is 'easy' to learn and affordable. There are alternatives to spreadsheets for medical statistical analysis such as SPSS, R and Matlab. Both R and Matlab resemble programming languages and require a detailed knowledge of the syntax and argument construction to be wielded effectively. SPSS requires less specialised knowledge and resembles a spreadsheet. However, SPSS costs significantly more per license than Microsoft Excel (SPSS is starts at $1170 USD per year as a subscription, Excel costs $331 for a full user owned copy). Between the usability, flexibility, perceived 'easiness' of spreadsheet software and license cost, it is no surprise that spreadsheet software is the de-facto choice for data analysis in the medical field. Indeed, it is these same reasons that spreadsheet software is used extensively in the business world.

**1.4 Spreadsheets Risks**

Spreadsheet software is amongst the most utilised commercial software in organisations world-wide. Users cite 'ease of use' and a wide range of functionality that could replace many complex information systems. The business world makes extensive use of spreadsheet software for data processing, analysis, decision science and data storing needs (SERP, 2006).

Spreadsheets are however, prone to multiple risks and since they are standalone files, they lack system-wide controls off the shelf. Almost any employee can create access, manipulate, and distribute spreadsheet data. Hence, almost any employee can make a small risky error while manually entering data or configuring formulas (Deloite, 2009).

Spreadsheets errors are prevalent and can cause crises in organisations. Spreadsheets error rates and cell error rates are high and there are many real stories worldwide that show that these errors can cause serious problems in the business world (EuSpRIG, 2016). Through many years of research, it has become clear that spreadsheet use can carry multiple serious risks (Panko, 2008).

The risk of making a simple mistake appears to be high, field studies show error rates shows that up to 90% of spreadsheet models contain at least one error (Panko, 2008). Spreadsheets are created and used without proper documentation and organisations generally do not have strict criteria governing their use.

Loss of data is a particularly dangerous risk having all data on spreadsheets and not having a centralized and data recovery environment could lead to a crisis in any financial or nonfinancial system. Data availability environments should be created also to ensure business continuity in the event of data loss.

Unskilled users could be considered as a risk since business spreadsheets are designed by both IS and non-IS professionals. The important issue to consider here is that non-IS professionals are unlikely to be trained in information systems development methods, meaning that the process of creating a spreadsheet is far more ad-hoc and is unlikely to follow standards dictated by software engineering. Indeed research shows that almost all spreadsheet modellers have no formal training (SERP, 2006). Because of this, it is impossible to guarantee the adherence of standards to any one spreadsheet modeller or the validity of a particular spreadsheet model. Research shows that most errors do not arise from mistakes in programming the spreadsheet, rather they arise from the misapplication of programming logic (Panko, 2008). This makes the lack of user training in formal development methods even more critical since, once committed, logic errors are difficult to find and correct. Without the knowledge of how to test and debug the spreadsheet, the chance of a user noticing and correcting such a mistake is low.

**1.5 The Pilot Study**

This paper considers the research environment in the neuroscience department at the University of Newcastle. Within this department, there is diverse research being undertaken ranging from the basic biology of neurons to the abnormal activity associated with epilepsy, from music perception to mood disorders, from visual object recognition to retinal prostheses for the blind, from animal decision-making to anaesthesia to neurological disease. The department has various tools at its disposal including:

- Brain scanning (MRI)
- Cellular imaging and electrophysiology
- Computational modeling
- Molecular genetics
- Animal and human behavioral laboratories
- Psychophysics

The neuroscience department consists of 17 researchers comprised undergraduate, postgraduate and post-doctoral students. There are also a number of senior academics within the department. The department produces world class research, publishing papers in leading neuroscience outlets such as the *Journal of Neuroscience, Brain and Language* and *Physics life reviews*.

**1.6 Research Materials and Methods**

A number of different research materials are employed to gather relevant information. Firstly an in-depth questionnaire across all members of the research centre was distributed. There are also two in-depth 45 minute interviews conducted with senior members of the department to supplement this information.

**1.6.1 Questionnaire**

The questionnaire was designed to gather information on several different dimensions of spreadsheet use in the Neuroscience lab. The questionnaire posed questions on the following themes: Participant demographics; The importance of research data; Spreadsheet use in the organisation; Participant knowledge and experience of spreadsheets; Spreadsheets and other statistics software; The spreadsheet lifecycle; Backup and security.

The questionnaire contained 35 questions and although the large majority were closed multiple choice questions there were also some open free typing questions for balance.

| Question area | Questions posed |
|---|---|
| Demographics | Age, Sex, Education and Role in the organisation |
| The Importance of research data | Frequency of spreadsheet use; size of spreadsheets in the department; number of users per spreadsheet and motivations for spreadsheet use. |
| Spreadsheet knowledge and experience | Methods of learning spreadsheets; self-assessed proficiency and willingness to train |
| Spreadsheets and other statistics software | How useful are spreadsheets for data analysis; other potential statistics software and personal advantages of spreadsheet software |
| The spreadsheet lifecycle | Approaches to design, separation of input, calculation and output; use of guidelines in development; approaches to testing; documentation |
| Spreadsheet backup and security | Organisational backup strategies; cell protection; password protection; |

Table 1 Questionnaire areas and questions

**1.6.2 Interviews**

Two semi-structured interviews were conducted with two post-doc researchers in the laboratory. Post-doc researchers were chosen since they were likely to have a more mature, fuller understanding of conducting research, analysis and the workings of the lab. In addition, the post-doctoral students are those who are still working in the lab, whereas the senior members of the academic staff assume a more supervisory role for the undergraduate and postgraduate students. Both interviews lasted around 45 minutes, a range of questions were posed to each interviewee, see table 2.

The interviews were designed to get the participants to extrapolate on the answers provided in the questionnaire and to explore the detail and subtlety of the relationship between spreadsheets and scientific research.

| **Questions relating to conducting research** | <ul><li>How many experiments do you conduct in the year?</li><li>How often do you deal with research data?</li><li>What type of data do you usually deal with in your research?</li><li>How many people work on the same experiment and share work together?</li></ul> |
|---|---|
| **Questions relating to spreadsheet use** | <ul><li>Do you use spreadsheets daily?</li><li>Why do you use spreadsheets?</li><li>What is the percentage of time using spreadsheets compared to other software?</li><li>Do you back up your spreadsheets? How often? What method?</li><li>Who can access your spreadsheet? Could he/she edit?</li><li>Are there any guidelines for using spreadsheets? Documentation?</li><li>How do you protect spreadsheets data?</li><li>What level of Knowledge do you have on spreadsheets?</li><li>Is there any second level of checking? What is the method (code inspection…etc.)?</li><li>Are there any preparation steps or guideline to create spreadsheets?</li><li>If data is lost what are the effects (in terms of effort, reputation, legislation and finance)?</li><li>If the data has errors, what are the arising issues? (in terms of effort, reputation, legislation and finance)?</li><li>If data is lost is it hard to get it again?</li><li>What is the period of time needed to collect research data?</li><li>Is there any policy regarding dealing with spreadsheets?</li><li>How much does it cost to run an experiment? (Per day and Per experiment)</li><li>How is your research funded?</li><li>Lack of training and overconfidence - are these a problem from your point of view?</li><li>Is there a risk of losing funds?</li><li>Is there a risk of delaying a research results?</li></ul> |

Table 2 Interview questions

### 1.6.3 Limitations

The number of respondents was relatively small with only 17 completing the questionnaire and two completing interviews. This obviously limits the generalisation that can be drawn from the research but it still provides an interesting initial view of the risks and difficulties of the use of spreadsheets in this clinical setting. There is very little research presently that considers the influence of electronic resources in clinical medicine and hence this paper presents a valid contribution to this field.

Proceedings of the EuSpRIG 2016 Conference "Spreadsheet Risk Management" ISBN : 978-1-905404-53-7
Copyright © 2016, EuSpRIG European Spreadsheet Risks Interest Group (www.eusprig.org) & the Author(s)

## 2.0 Results

The analysis identifies spreadsheet risks in the following themes: Training; Overconfidence; Spreadsheet design approaches; Documentation.

### 2.1 Training

The results of questions related to training shows that none of the respondents held any certified training in spreadsheets, see figure 1. More than two-thirds of the respondents learned spreadsheets through self-tutoring. This lack of training is typical finding of almost all surveys of spreadsheet training practice. (Taylor *et al* 1998, SERP 2006). Self-tutoring does not necessarily indicate a lack of understanding but it does likely suggest that those participants are not formally trained in approaches to planning software, managing data and objectively testing for errors. Spreadsheet modelling has been identified as a cognitively complex activity comparable to that of medical diagnosis (Kruck, 2003). However, consider that routine medical diagnosis is only possible after years of specialised training. Superficially, spreadsheets seem to be simple and straightforward tools but error rates show they carry significant risk. Without an understanding of approaches to planning, development and testing, it would seem more likely that mistakes committed would go unnoticed.

However, a lack of training is not considered an unusual situation as one of the participants suggested in the interview:

> "In research, you just have to pick new skills as you go in job. So, peer learning and picking up things by reading online manual is something all researchers have to do".

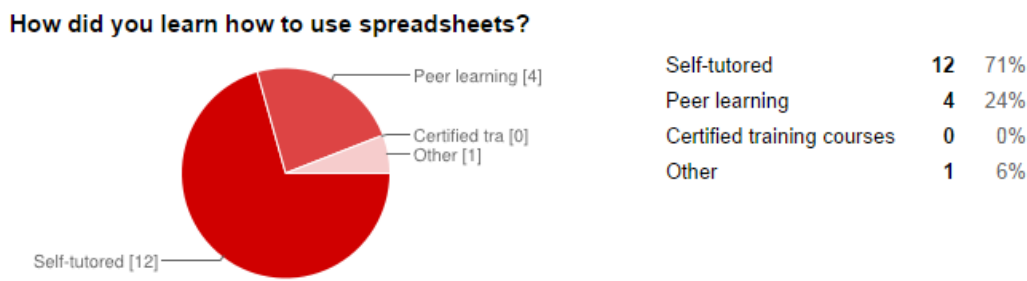

**Figure 1 Spreadsheet training**

According to the 'internal auditor' (Larry, 2007) a lack of training is identified as the first source of risk when dealing with spreadsheets. Larry (2007) recommends that 'untrained users' should be educated to more than a basic level to reduce the risk of error.

### 2.2 Overconfidence

Ease of use and overconfidence are related phenomena. The questionnaire showed that although none of the respondents had a certified training, more than three quarters claimed that they are on or above the intermediate level, see figure 2. Overconfidence tends to increase specially in people who are highly educated and doctoral research represents the highest level of qualification in the education system.

Formal training in spreadsheets is something the researchers do not look at as a necessity, as one of the interviewees replied. It might be correct that all sources of knowledge are available online, but there are a lot of hidden risks that could be eliminated by simple steps. For example, multi-level code inspection could eliminate errors significantly, but it is not something that is typically recommended in online instructional material.

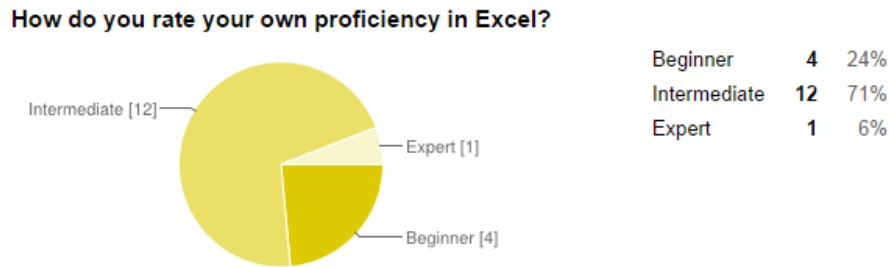

Figure 2 - Proficiency in spreadsheets

Overconfidence is risk as discussed by Panko (2008) said: *"Overconfidence is corrosive because it tends to blind people to the need for taking steps to reduce risks"*. In his paper, Panko realized the severity of this issue and the fact that it exists more than we think which is reflected in figure 2, and is referred to in his previous research:

> *"… when Brown and Gould gave three spreadsheet development tasks to nine highly experienced spreadsheet developers, all made at least one error, and 63% of the spreadsheets overall had errors. Yet when asked about their confidence in the correctness of their spreadsheets, their median score was "very confident."*

Overconfidence affects both novice and expert users (Panko, 2008). It also appears that the self-efficacy of the individual modeller is the most important factor when considering how overconfidence can be observed and mitigated (Takaki, 2005).

## 2. Design and testing

Almost all of the researchers (94%) start designing their own spreadsheets by directly inputting data into the computer. This finding is typical of most spreadsheet user surveys (SERP 2006, Taylor *et al.* 1998) and highlights a particularly risky behaviour amongst most spreadsheet developers. Several spreadsheet design standards recommend sketching the design on a piece of paper first (Grossman 2002, Grossman and Ozluk, 2004). However, none of the respondents indicated they sketch the model on paper first, instead they go straight to the spreadsheet and enter formulae and data directly. It would also seem that such behaviour is consistent with overconfident modellers, as indicated in figure 2 – i.e. an overconfident modeller does not see the need to plan the model before they start coding.

When asked how they approach testing the spreadsheets they create, the largest majority (40%) indicated they use self-revision checks to ensure the model is free of errors, 13.3% indicated they did not use checks, 20% said they submit spreadsheets for peer review and 26.7% said they use a calculator to check the accuracy of the calculations, see figure 3. This is an encouraging finding since most participants do some form of checking. The most promising of these are using a calculator and peer review. Peer review is a good idea from the perspective that if the work contains errors, it is more likely that the error will be spotted by a team than an individual. Research shows that spreadsheets audited by teams generally find more mistakes, in one study (Panko, 2008) individuals found and corrected 60% of an error seeded model, whereas groups of three found and corrected 80%. The use of a calculator to check the accuracy of the mathematics in the spreadsheet is a promising approach since it requires the tester to reconsider the dimensions of the model, indeed one would need to examine all assumptions to effectively check using this approach. If there had been a mistake in the logic or mechanics of the spreadsheet, cross checking the outcome with a calculator might expose those mistakes (Colver 2008, O'Beirne 2009, Ayalew *et al.* 2000).

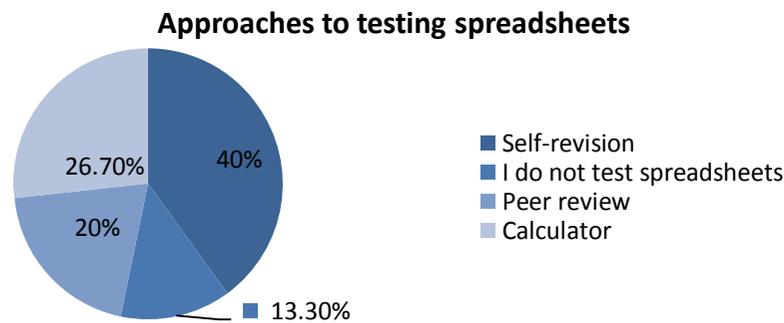

Figure 3 Approaches to testing spreadsheets

### 2.4 Documentation

Documentation is an essential aspect of the spreadsheet life cycle. Ideally, every spreadsheet should have supporting documentation to describe the input, computation, output and data points contained in it.

Documentation assures better understanding of the spreadsheet (Pryor, 2004) which leads to better accuracy and fewer errors. In addition, it is particularly critical when there are multiple users using the same spreadsheet. The participants indicated around a third of the respondents share spreadsheets with other users. The questionnaire showed that on average, 2-5 researchers share the same spreadsheet.

The data also showed that nearly a third of the respondents do not document their spreadsheets at all. And nearly half of the respondents use the cell-comments feature to document. It is encouraging that comments feature at all since this can be viewed as a kind of documentation. However, cell comments can be thought of as the equivalent as annotated code in software engineering, whilst these comments are important and assist in making the spreadsheet more understandable, it does not offer the depth of analysis or guidance as a conceptual model might.

### 2.5 Security

This research considers the spreadsheet security from several different angles that align with the Spreadsheets Standards Review Board (SSRB) standards (SSRB, 2003). The first line of protection is setting a strong password on the spreadsheet. The majority of the respondents (82%) didn't use this feature to protect their spreadsheets. Moreover, the majority that used passwords had them written down somewhere in the office. Finally, cells that contained output data had not been protected from manipulation. Since more than third of the respondents share spreadsheets, the output cells should be protected to prevent accidental overwriting of formulae.

However, the IT infrastructure of the premises offsets the risk of theft via local secure storage. Each lab has its own share drive to store spreadsheets and data and only. Access is only granted to the lab researchers so the chance of theft is greatly reduced. This does not mitigate the risk of accidentally overwriting cells in the spreadsheet.

### 2.6 Backup solutions

The backup 'rule of three' states that for a file to be sufficiently backed up it should be kept in three separate locations using two different types of media with one offsite backup.

A lack of an adequate backup solution could mean permanently lost data, effort and time. In this research, more than 82% of the respondents seemed to be unaware of suitable backup procedures to protect their data. Some respondents kept a single backup of work on external hard disks. Others used the Universities local networked servers as their means of backup. Whilst the networked infrastructure of the university offers some security from lost files, it does not meet the conditions of the 'rule of three' and hence one should not consider these files adequately backed up.

During the interview stage of the study, interviewees were asked about the consequences of losing research data:

> *"Losing data could result in a lot of time and effort being put in to repeating the research if results were not also recorded elsewhere (which they usually are). If repeating experiments is necessary this may cost the organization to fund another set of experiments. This would also delay publishing the data which could lead to a delay in publishing (manuscripts) which would also result in a delay in any benefits to the general population the research may provide. For example the creation of new treatments"*

> *"...damage to your personal reputation, huge amount of time wasted and have to do it again, would have to prove to tutor/supervisor that it won't happen again somehow which is likely really hard in today's dog-eat-dog research society Organization. Ethical consequences of not keeping original results, would practically invalidate studies that might have been published, if other researchers/ethics committees wanted to look at original results we would be in legal trouble, staff members may be sacked, even criminally liable in some cases. Media coverage of bad science would reflect badly on the university…"*

> *"lost effort collecting data, lost potential for important discoveries which could have had widespread implications"*

Clearly the risks of losing research data are severe to the participants, the university, the research discipline and the wider public. The questionnaire data suggests that whilst some backup precautions are followed, there is more that could be done to secure the departments data.

### 2.7 Complexity, Frequency of use and number of users

Defining complexity in spreadsheets objectively is difficult since the scope and use of functions varies from one application to another. Therefore, data on the type of features used in spreadsheets was used to determine its complexity in addition to determining the percentage of cells with formulae, see figures 4 and 5. This is broadly in line with other studies of functionality use (Chan and Storey 1996, Ballinger *et al.* 2003, Thorne and Ball 2008)

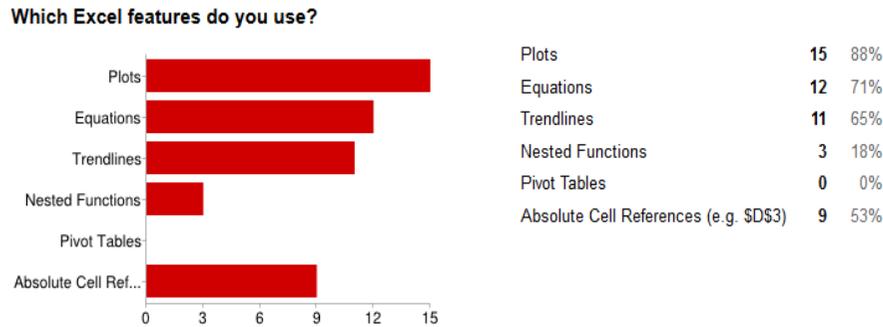

Figure 4 Excel features used

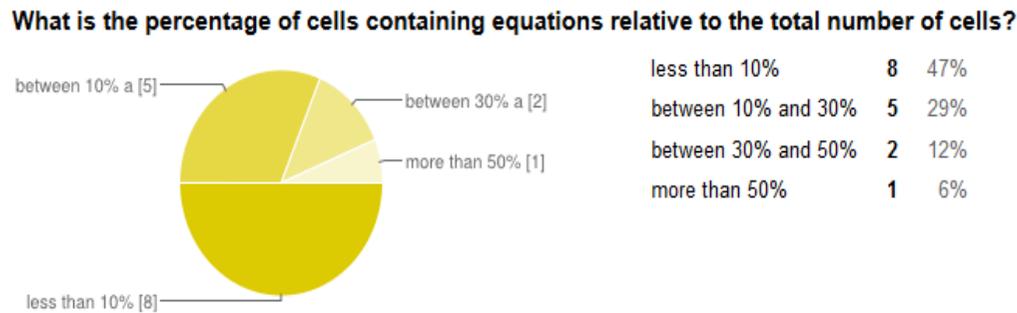

Figure 5 Cells containing equations

From figures 4 and 5, it is obvious that the majority of the researchers use spreadsheets to plot data and do some calculations using equations. However, around half of the respondents indicated that less than 10% of their spreadsheet cells contain formulae. The majority of the participants indicated they use 'simpler' excel functions in their spreadsheets rather than make use of nested functions which are more complex to program. It would seem then that the complexity of the spreadsheets at the lab is at the lower end of the complexity spectrum. This picture is fairly typical

### 2.9 Spreadsheet errors impacts

This section discuss some of the potential impacts arising from the use of spreadsheets in neuroscience research. The answers in this discussion arose mostly from the interview process and examine a specific set of risks for those working with scientific research data.

### 2.9.1 Financial loss

Although financial issues are well considered in business data, financial loss should be taken into consideration when dealing with research data.

The neuroscience lab at Newcastle University is funded through a mix of European Science Council funding, internal funding and

Establishing a science lab, such as the one study, with state of the art technology is a costly exercise. The lab also has significant 'on-costs', meaning that the running costs sometimes exceed the setup. Every experiment run in the lab therefore has its own cost which is deducted from a limited budget. According to the respondents, the average cost per neuroscience experiment is around £1000 per day. However, the cost varies according to the type of experiment, the tools used, the materials available and the number of staff involved. Since the costs are so high, any data generated from these experiments is valuable, and losing data would mean a significant financial expenditure to re-gather the data. Furthermore, the budgets are tight and there is no space for errors or time wasting.

If data was lost, additional funds to rerun the experiments might be sought as one of the respondents described:

> *"Lots of effort would be required to correct the data by, for example, rerunning that part of the experiment. This would also be likely to cost more money and might require additional legislation"*

To conclude, financial loss is critical to the organisation, but it would even be catastrophic if financial loss as combined with other impacts.

**2.9.2 Wasted effort and data loss**

Wasted effort is one of the key impacts that almost all of the participants agreed on. Wasted effort includes the effort of the original experiment, the actual data collection, error checking and the effort to correct or re-run the experiment.

> *"Lots of effort would be required to correct the data by, for example, rerunning that part of the experiment"*.

> *"It takes a long time to go back and check spreadsheets for errors"*

> *"The biggest effects would be loss of effort and reputation"*

> *"It would take a long time to possibly correct the errors, or if the results hadn't been backed up elsewhere, the research would have to be carried out again"*

> *"The biggest impact from input errors would be the time consumption of checking for errors"*

Clearly the participants recognise the risks of having to re-run the experiments and the significant monetary costs of repeating experimentation. A single error that materially affects the outcome of the experiment would cost a significant amount of money to correct. This is especially true when the data collection period lasts months.

In addition to the wasted time, wasted data is also crucial. More than two-thirds of the respondents of the questionnaire selected *"Losing valuable experimental data or simulation results"* as one of the impacts and risks of spreadsheets errors.

### 2.9.3 Lower Productivity

Each research group in the neuroscience institute has between 18 - 20 experiments per year to conduct. Therefore, having errors and wasting time could reduce this number which in turn would lead to a significant reduction in the productivity of this group in terms of research conducted, outputs produced (Academic papers) which in turn has a knock on effect on the credibility of the department and the University.

> *"It may reduce the amount of papers I produce in my PhD therefore reducing my attractiveness to a future employer"*.

### 2.9.4 Causing unnecessary suffering to animals

One of the respondents pointed out that unnecessary animals suffering as one of the impacts of errors in research data. Newcastle neuroscience department uses animals in some of their experimentation. Therefore, losing data or even having errors in it not only causes loss of effort and money, but also wasted days of experiments on animals.

No researcher or scientist intends to harm or cause suffering to animals if it could be avoided. Therefore, there is a significant ethical impetus to handle experimental data correctly and carefully.

### 2.9.5 Publishing corrections in publications

In the worst case scenarios, data with errors could pass all internal checks unnoticed and get published. In this case, the honesty of the researcher plays a major role in not misleading the discipline, journal and readership. One of the interviewers acknowledged the impact of errors in published papers, and when he was asked about the steps he would do if he discovered any errors in his experiments: *"You have to be very honest if you find any mistake and to make sure your correction reach everybody read your publications"*.

However, the risk of an error reaching the publication stage is lower since the paper will go through at least one process of peer review when it is being considered for publication. A paper is also likely to be reviewed internally, especially if the paper is the result of PhD study or research grant further reducing the chance of the paper reaching publication with an error. Multiple authoring further reduces the likelihood of unnoticed mistakes since the work is likely to be read and considered by multiple experts.

> *"It is unlikely results with errors would be published into the scientific community since there are many stages where different people check and analyse the results. In summary: If there are errors more time and effort would be used"*

### 2.9.6 Unreliable research

Although internal and external peer review processes are likely to highlight errors, there is a chance that manuscripts with errors could be published in a journal. This could mean that research is published with misleading conclusions. A good example of this is the Reinhardt and Rogoff paper

"Growth in a Time of Debt" which was published in the peer reviewed journal *American Economic Review* in 2010. This paper contained a statistical analysis that suggested that if a countries external debt exceeded 90% of the country's Gross Domestic Product (GDP), negative growth was the consequence. However, the statistical analysis contained a flaw which meant that the 90% figure was incorrect. In actual fact, the analysis was completely wrong, a corrected analysis issued by Herdon *et al.* 2013 showed that when debt exceeds 90% of a country's GDP growth is still positive. In the time between the article being published and the discovery of the error, the paper was used by a number of governments in the US and Europe as a justification for austerity. The paper was even quoted directly by the UK chancellor George Osborne in his discussions on the economy. Hence decisions were made and policies formed on erroneous data.

Neuroscience is a complex field of study with only a few institutions globally dedicated to neuroscience research. It is conceivable that an error could reach publication and that the published manuscript could form the basis for real world decision making as in the Reinhardt and Rogoff case.

Unlike books, research publications in conferences and journals do not generally have post-publication review and feedback. This ensures fewer corrections to the journal, but it also means that if a mistake makes it through the peer review process that the work is unlikely to be questioned.

When asked about the chance of errors creeping into published manuscripts, the participants recognised the risk of unreliable research:

> *"Unfortunately, since the data are published there is no feedback and the data are considered as right"*

> *"In general, making errors in the data involve having wrong conclusions and guiding the researcher to the wrong hypothesis"*

### 2.9.7 Misleading science

Errors in research data have severe impacts on the science fields in general. Although the consequences of an error could include financial loss, some of the potential impacts could carry far greater ramifications. Unlike Reinhardt and Rogoff, some neuroscience researchers are dealing with real lives and the impact of an error could be catastrophic to the health of individuals following treatment plans informed by such research. Again, the participants recognise these issues in both the questionnaire data and follow up interviews.

*"Drawing incorrect conclusions from data leading to negative research impact"* was one of the multiple choice answers to the impacts and risks of errors in spreadsheets. 14 respondents out of 17 selected this choice in the questionnaire. Therefore, this issue is clearly recognised as a significant risk in scientific research.

One participant distinguished the impact of erroneous data based on the stage of its discovery:

> *"If the error is detected before publication, only the lab wastes time/effort/money. If the error is not detected until after publication, the lab/university's reputation will be affected. If the error is not detected at all, the entire field is misled"*

**2.9.8 Damage to Reputation**

Respectable research institutions are expected to publish reliable and well explored research manuscripts. Errors that reach publication could therefore severely damage the reputation of the institution if the research was widely published as many of respondents described:

> *"It is damaging to reputation to retract any statements/findings"*
>
> *"The organization would lose credibility because of poor research techniques, which would impact the university as a whole"*
>
> *"The reputation of the organization would be negatively affected if it turned out that errors had arisen in the data and not been checked/realized."*
>
> *"It is demoralizing, bad for reputation of the researcher and the school, time consuming"*

Other respondent highlighted a case that occurred in the medical research field recently by saying:

> *"if the errors are highlighted as it was seen for a recent Nature paper published by Japanese students, that the reputation of the lab considerably decrease and it may happen that some people lose their position"*.

**3.0 Conclusions**

The following conclusions are divided into three sections: The risk profile of the participants; The risk profile of the spreadsheet artefacts and the impacts of risk. Finally a comparison is made between business and academic spreadsheet risks.

**3.1 Risk profile of the participants**

Based on the questionnaire and interview data, the profile of the participants of the study are typical of similar studies in business research. Almost all participants indicated that they had no formal training in spreadsheet software, which is typical of spreadsheet users irrespective of the discipline of the sample. The participants also displayed classic evidence of overconfidence in the answers to a number of questions. When asked to evaluate their own competence in spreadsheet development, the large majority of the participants (76%) chose to rank themselves as "intermediate or expert" despite not a single participant having received any formal or certified training. When asked about spreadsheet design, 94% of the participants indicated that when developing spreadsheets, they opt to directly enter data and formulae into the spreadsheet without planning the model first. When developing spreadsheet models the participants produced very little supporting documentation, although some participants used cell comments to document certain features. Very few participants (18%) chose to protect their spreadsheets with cell locking on the sheets. The most significant reason to lock the cells on a spreadsheet is to prevent yourself or others accidently overwriting portions of the spreadsheet, this practice particularly prudent if the spreadsheet is being shared with others.

**3.2 Risk profile of spreadsheet artefacts**

Based on the information gathered from the participants, the level of complexity assigned to the artefacts is intermediate. Figures 4 and 5 show a 'simple' use of spreadsheet functions and relatively small spreadsheets. However, when looking at research outputs of the department, for instance (Gregson *et al*, 2014), it is clear that statistical analysis techniques are utilised to validate data. Presumably this analysis is done within Excel and therefore could redefine the complexity classification as high. However, this is anecdotal since none of participants indicated they perform these relatively complex calculations in the answers given. The questionnaire data also suggested that the number of new spreadsheets being created is relatively low, with participants indicating they tend to work on the same spreadsheet frequently.

### 3.3 Risk impacts in research

The specific risks arising from this use are numerous and will be listed sequentially in order of potential negative impact to the organisation**.**

The first potential risk is financial loss through lost data. The neuroscience lab has significant running costs, these costs are a combination of staffing, equipment and materials for experimentation. The average cost for running an experiment in the lab is around £1000 pound a day. Hence if spreadsheet data was lost through whatever means, it would cost a significant amount of money to re-run the experiments which would have to be accounted for as part of the Universities tight budget for running this lab.

Wasted effort and lost data are both significant risks to the researchers at the neuroscience lab. Any lost data would take significant time and effort to reproduce. In the lab environment, the amount of time available for the researchers to use the lab is limited by budget and demand, meaning that finding spare capacity and budget to re gather data is difficult.

Having to repeat experimentation also means that the overall productivity of the lab and the researchers is impacted since fewer measurable research outputs (such as academic journal papers) can be published if experiments must be re-run.

Since some of the experimentation is done with live animals, any errors or lost data could mean that any suffering the animals experienced as part of the research was unnecessary. Obviously this is of the upmost ethical importance to the staff who would never intentionally cause suffering to animals. Indeed, if it was possible to do the experimentation without the animal subjects then that would be the de facto position however it remains the case that many medical breakthroughs have been made through the use of animal subjects.

Errors in data that reach publication could have a number of serious consequences for the institution. Firstly errors in data or interpretation of data would equate to published research that would be considered unreliable. It's possible that others could then make decisions on the basis of erroneous research such as the Reinhardt and Rogoff case, the consequences here are that the conclusions are misleading. This may in turn influence other researchers in to making poor decisions, such as choosing research questions based up erroneous conclusions.

### 3.4 Comparing academic and business risks of spreadsheets

The profile of spreadsheet modellers between the academic and business world is similar. This is evidenced through a lack of formal training, issues of overconfidence, a lack of preparation and documentation. These findings are typical of spreadsheet surveys conducted in the business environment too, hence the profile of spreadsheet modellers in academic and business worlds are similar.

One of the major differences between these worlds are the mechanisms that each uses to communicate information externally. In the business world, information contained in spreadsheets might be communicated through a variety of means to the general public but importantly, it is up to the business to ensure that the information being communicated is sound. In the academic community, most external communications comes through research outputs such as papers published in conference or journal proceedings. Typically, these research outputs are double blind peer reviewed before publication. This provides academic research with a significant safety net when considering spreadsheet errors, since if mistakes are made in publications or if falsified data is presented, it is likely that these issues will be spotted before being communicated to the community. Of course such processes aren't infallible (Reardon & Cyranoski, 2014) but they must prevent a good proportion of mistakes from reaching the public eye. Peer review processes have long been advocated as one of the best means to reduce the likelihood of spreadsheet errors (Panko, 2008) – this pilot study really underlines that issue since the number of retractions in the academic world are very low. Hence it would seem that the peer review process is particularly effective at preventing spreadsheet errors (amongst other types of errors) from reaching publication.

**3.5 Identified areas for future research**

This study highlights two main opportunities for further interesting research, the first considers the internal peer review process and the second focuses on the 'important' spreadsheets in the organisation. To that end, the following research questions have been identified:

1. How are processes of internal verification and review implemented at the Newcastle neuroscience lab to reduce errors in spreadsheets, research and publication?

This research question will explore how the internal peer review process is used to reduce defects in spreadsheets, research and publications. A successful system of this type could provide a model approach that could be adopted by other organizations to reduce the likelihood of spreadsheet defects be communicated outside of the organization.

2. What is the character and make up of 'important' spreadsheets in the neuroscience lab and what specific risks do these spreadsheets carry?

This research question will offer a detailed examination of spreadsheets identified as 'important' to the lab. The spreadsheets will be examined in terms of programming structure, functions used, process of development and documentation. The spreadsheets will be analysed for vulnerabilities and 'what-if' risk scenarios will be explored with the modelers.

**3.6 Conclusion**

This pilot study has examined spreadsheet use at the neuroscience lab at the University of Newcastle. From the study we can conclude that spreadsheets play an important role in conducting, recording, analysing and communicating scientific research. The risks present in these activities relate to misanalysis, misrepresentation, data loss or corruption, wasted time, lower productivity and unnecessary suffering to animals. The potential consequences of such risks mainly relate to the publication of data or analysis that is erroneous or misleading. However, the neuroscience lab has a process of internal verification that explores the data, analysis and inference before they leave the confines of the laboratory.  A more detailed examination of the 'important' spreadsheets and the internal verification process is proposed since a process like this could be adapted to a business context to help reduce the number of errors in the business world.